\documentclass[elsarticle]{revtex4-2}

\usepackage{xcolor}
\usepackage{graphicx} 
\usepackage{subfigure}
\usepackage{dcolumn} 
\usepackage{bm}
\usepackage{float}
\usepackage{hyperref}
\usepackage{amsmath}
\usepackage{lineno} 

\begin{document}

\title{Probing Spin and Lifetime Correlations in Entangled Hyperon-AntiHyperon Pairs}
\author{Aihong Tang}
\affiliation{Brookhaven National Laboratory, Upton, New York 11973, USA}
\date{\today}

\begin{abstract}

Quantum entanglement has now been demonstrated in several hadronic systems, revealing that non‑classical spin correlations survive even through the strong‑interaction hadronization process. To date, however, all studies have focused exclusively on angular observables, leaving the possibility untouched that quantum coherence might also influence the decay times of entangled partners. In this work we propose data-driven tests of spin-lifetime and lifetime-lifetime correlations for $\Lambda–\bar{\Lambda}$ pairs produced in high‑energy collisions. By examining the opening‑angle distribution in slices of $\Delta t$, constructing a pair-wise spin-lifetime correlator, and testing a simple lifetime-lifetime covariance, we search for deviations from independent exponential decay that align with known spin correlations. Observation of nonzero lifetime correlations would compel a reassessment of how entanglement manifests in decaying systems, revealing hitherto unexplored temporal coherence.

\end{abstract}

\maketitle

\section{Introduction}

The phenomenon of quantum entanglement, popularized by the landmark optical-photon Bell tests~\cite{PhysRevLett.28.938,PhysRevLett.49.91}, has also been observed in hadronic systems, for example through the polarization correlations of $\Lambda\!-\!\bar{\Lambda}$ pairs produced in $e^+e^- \rightarrow J/\psi \rightarrow \Lambda\bar{\Lambda}$ at BESIII~\cite{BESIII:2019epr,Wu:2024asu,Fabbrichesi:2024rec}.  Entanglement measurements have further served as powerful probes of quantum correlations in high-energy and nuclear experiments.  In particular, over the past few decades, time-dependent interference in neutral-meson systems has been explored in a variety of contexts—from early fixed-target and collider measurements~\cite{ARGUS:1987xtv,CPLEAR:1998dvs}, through $B$-factory studies~\cite{BaBar:2001pki, Belle:2001zzw,Belle:2002ghj,Belle:2007ocp}, to the $\phi$-factory kaon-pair interference~\cite{KLOE:2006iuj}—revealing unmistakable $\Delta t$-oscillations in $K^0\!-\!\bar{K}^0$ and $B^0\!-\!\bar{B}^0$ pairs.  Outside meson systems, violation of the Leggett–Garg inequality has been observed in neutrino oscillations~\cite{Formaggio:2016cuh}.  More recently, the ATLAS and CMS Collaborations extended entanglement tests to the quark level by observing a $>5\sigma$ spin-entanglement signal in $t\!-\!\bar{t}$ pairs produced in $pp$ collisions at $\sqrt{s}=13\,\mathrm{TeV}$~\cite{ATLAS:2023fsd,CMS:2024pts}, and the STAR Collaboration demonstrated that, via $s\!-\!\bar{s}$ quark pairs, quark-level entanglement survives nonperturbative hadronization by measuring significant spin correlations in $\Lambda\!-\!\bar{\Lambda}$ hyperon pairs~\cite{STAR:2025njp}.  In addition, the BESIII Collaboration has exploited hyperon spin entanglement for precision tests of CP symmetry~\cite{BESIII:2021ypr} and local realism~\cite{BESIII:2025vsr}.

Motivated by the broader lineage of entanglement studies in neutral-meson and neutrino systems~\cite{KLOE:2006iuj,Belle:2007ocp,Formaggio:2016cuh}, we propose and develop data-driven probes of spin and lifetime correlations in entangled \(\Lambda\!-\!\bar{\Lambda}\) pairs produced in high-energy collisions. In our approach, both the simple lifetime–lifetime covariance and the \(\Delta t\)-dependent angular analyses (together with a per-pair correlator) are used to uncover genuine \emph{time-domain entanglement}—by which we mean nonclassical correlations that appear directly in the decay times of spatially separated partners.  Unlike neutral-meson \(\Delta t\) oscillations, which arise from mass-eigenstate mixing, and unlike conventional spin-correlation measurements, our observables directly test nonfactorizability in the temporal decay variables themselves, without assuming a prior operator representation for lifetime.  While decay time is not a standard Hermitian observable, each decay acts as a generalized quantum measurement on its parent~\cite{Shi:2019kjf,Qian:2020ini}, and nonfactorizability of the joint wavefunction can induce correlations among these measurement outcomes.  For brevity we use the term “entanglement,” but more generally the proposed observables test for quantum coherence in the temporal domain arising from a shared, nonseparable state.  These constructions are model-independent in their null hypothesis and simultaneously serve as sensitive discovery tools for richer, spin-modulated temporal structure.  Observation of nonzero lifetime correlations would compel a reassessment of the standard framework for entanglement in decaying systems, revealing that coherence can persist and manifest in the temporal domain in ways not previously accounted for.

A particle’s decay can be regarded as the result of a generalized quantum measurement.  In previous treatments, the decays of $\Lambda$ and $\bar{\Lambda}$ in an entangled pair were regarded as independent measurements~\cite{Wu:2024mtj}, implying no intrinsic correlation in their decay times.  Importantly, however, no known symmetry principle or superselection rule in quantum mechanics forbids correlations between the decay times of two spatially separated entangled particles.  Once two particles are entangled they share a single nonseparable wavefunction, so a measurement that partially collapses it on one side constrains the remaining degrees of freedom on the other—not by signal propagation, but because the joint state cannot simultaneously occupy incompatible branches.  By contrast, for truly independent decays each hyperon’s survival probability follows its own exponential law, so the joint probability factorizes as
\begin{eqnarray}
P(t_1, t_2) = P(t_1)\,P(t_2)
\end{eqnarray}
and any covariance \(\langle (t_1 - \bar{t})(t_2-\bar{t})\rangle\) vanishes.  Furthermore, when viewed in isolation each decay time reflects the intrinsic memoryless stochastic nature of an unstable state, yielding the familiar exponential distribution: if \(P(t)\) is the survival probability and the instantaneous decay rate is constant \(\Gamma\), then
\begin{equation}
\begin{aligned}
P(t+\mathrm{d}t) &= P(t)\bigl(1-\Gamma\,\mathrm{d}t\bigr)\\
\Longrightarrow\,\,\, \frac{\mathrm{d}P}{\mathrm{d}t} &= -\Gamma\,P(t)\\
\Longrightarrow\, P(t) &= e^{-\Gamma t}\,,
\end{aligned}
\end{equation}
so the decay-time density \(f(t)=\Gamma e^{-\Gamma t}\) is purely exponential and two independent decays carry zero covariance. This factorized, independent-decay picture serves as our null hypothesis.

Before turning to the entangled case, we clarify what we mean by ``observing the $\Lambda$". Although charged particles can be tracked prior to decay, the neutral $\Lambda$ leaves no pre-decay track; In high-energy collision experiments, its decay vertex is reconstructed from the charged daughters, and its flight path is inferred a posteriori from the line connecting the primary and decay vertices. For brevity, when we speak of “observing the $\Lambda$ track,” we mean this vertex-to-vertex reconstruction. In general, such tracking constitutes a sequence of null-result checks (“still alive at $t'$”) that updates the state on survival but do not project onto a definite decay time; the decay itself is the temporal measurement that registers “decayed at $t$.” Position localization by tracking is a local operation on the spatial part of the state and does not, by itself, force factorization in other degrees of freedom (spin or temporal)—as in optical-photon Bell tests, where detectors localize particles while polarization entanglement persists~\cite{PhysRevLett.28.938,PhysRevLett.49.91}. With this operational point clarified, we now consider the entangled pair.

Accordingly, for an entangled \(\Lambda\!-\!\bar{\Lambda}\) pair, the survival amplitude
\begin{eqnarray}
A(t_1, t_2) = \bigl\langle \Psi \bigr|\,e^{-i H_1 t_1}\,e^{-i H_2 t_2}\,\bigl|\Psi\bigr\rangle
\end{eqnarray}
does not in general decompose into separate single-particle amplitudes, and the resulting joint decay-time distribution
\begin{eqnarray}
P(t_1, t_2) = \bigl|A(t_1, t_2)\bigr|^2
\end{eqnarray}
can therefore exhibit genuine correlations without violating any principle of quantum mechanics. In what follows we take the factorized, independent-decay form \(P(t_1,t_2)=P(t_1)P(t_2)\) as the null hypothesis and test for any departure from it.  Demonstrating a nonzero lifetime–lifetime covariance in the \(\Lambda\!-\!\bar{\Lambda}\) system would reveal a new facet of quantum coherence, enable direct tests of time-domain entanglement in unstable hadrons, and provide empirical constraints on decoherence. In particular, studying decay-time correlations directly in the lab frame—where moving hyperons experience time dilation—tests whether entanglement persists when lifetimes are relativistically stretched. This time-domain approach complements traditional angular measurements by incorporating kinematic effects into the decay-time observables.

\vspace{-1ex}

\section{Methodology}

We begin by recalling the established procedure for measuring spin–spin correlations in entangled $\Lambda$–$\bar\Lambda$ pairs \cite{PhysRev.186.1392, Tornqvist:1980af, Tornqvist:1986pe, STAR:2025njp}.  After acceptance corrections, the distribution of the opening angle $\theta^*$ between the decay proton and antiproton in their respective parent rest frames is fitted to
\begin{equation}
  \frac{1}{N}\,\frac{dN}{d\cos\theta^*}
  = \frac{1}{2}\bigl[\,1 + (\alpha_1\alpha_2\,P)\,\cos\theta^*\bigr],
  \label{eq:spinFit}
\end{equation}
where $N$ is the total number of same‐event pairs, $\alpha_{1,2}$ are the weak‐decay parameters, and the fitted coefficient $P$ quantifies the spin correlation.

To search for spin–lifetime correlations, we extend this framework by adding decay‐time information.  For each same‐event $\Lambda$–$\bar\Lambda$ pair we measure the decay times $t_1$ and $t_2$ (computed from the reconstructed decay length and momentum, $t = L\,m_\Lambda/(|\mathbf p|c)$ (proper time)) and reuse the same opening angle $\theta^*$.  We then form the raw per‐pair quantity
\begin{equation}
  w_i \;=\;\alpha_1\,\alpha_2\,\cos\theta^*_i,
  \label{eq:rawWeight}
\end{equation}
whose ensemble average in an uncorrelated (isotropic) sample is
$\langle w\rangle_{\rm iso}=0$ and variance $\langle w^2\rangle_{\rm iso}=(\alpha_1\alpha_2)^2/3$.  

Detector acceptance and efficiency effects are removed via a mixed‐event (ME) baseline: pairing each $\Lambda$ with a $\bar\Lambda$ from a different event matched in centrality and vertex position.  Denoting by $N_{\rm SE}(\cos\theta^*,\Delta t)$ and $N_{\rm ME}(\cos\theta^*,\Delta t)$ the counts in same‐event (SE) and mixed‐event samples for given $\cos\theta^*$ and $\Delta t\equiv t_2-t_1$, we define
\begin{equation}
  R(\cos\theta^*,\Delta t)
  = \frac{N_{\rm SE}(\cos\theta^*,\Delta t)}
         {N_{\rm ME}(\cos\theta^*,\Delta t)},
  \label{eq:jointDist}
\end{equation}
the acceptance‐corrected joint distribution in spin and relative decay time.  

Entangled lifetime correlations could appear either as a dependence on the time difference $\Delta t$ or as a more general covariance in $(t_1,t_2)$.  A test of $R(\cos\theta^*,\Delta t)$ projected onto $\Delta t$ alone probes the \emph{relative‑time} (marginal) distribution
\begin{equation}
  P(\Delta t) = \int_{-1}^{1} R(\cos\theta^*,\Delta t)\,d\cos\theta^*,
\end{equation}
but is blind to any \emph{common‑mode} shift of $t_1$ and $t_2$.  Conversely, an \emph{absolute‑time} test on the full joint distribution
\begin{equation}
  P(t_1,t_2) \propto R\bigl(\cos\theta^*,t_2-t_1\bigr)
\end{equation}
can reveal both $\Delta t$‑dependent and sum‑dependent correlations.  We therefore implement two complementary, data‑driven tests:

\subsection{Relative‐lifetime (\(\Delta t\)‐binned) test}

We partition the range of \(\Delta t\) into \(K\) bins centered at \(\Delta t_k\).  In each slice we extract the acceptance‐corrected angular distribution
\begin{eqnarray}
  R_k(\cos\theta^*) \;\equiv\; R\bigl(\cos\theta^*,\Delta t_k\bigr)
\end{eqnarray}
We then fit
\begin{eqnarray}
 r_k(\cos\theta^*) \;\propto\; 1 + P(\Delta t_k)\,\alpha_1\alpha_2\,\cos\theta^*
\end{eqnarray}
to extract \(P(\Delta t_k)\) with uncertainty \(\sigma_k\).  Under independent exponential decays $P(\Delta t)$ is constant;  we quantify any \(\Delta t\)‐dependence by
\begin{eqnarray}
  \hspace{-1cm} \chi^2 = \sum_{k=1}^K \frac{\bigl[P(\Delta t_k)-\bar P\bigr]^2}{\sigma_k^2},
  \bar P = \frac{\sum_{k=1}^K P(\Delta t_k)/\sigma_k^2}{\sum_{k=1}^K1/\sigma_k^2},
\end{eqnarray}
and compare to a \(\chi^2_{K-1}\) distribution.  A small \(p\)‑value signals a non‑flat \(P(\Delta t)\) and hence a relative‐time spin–lifetime correlation.

\subsection{Absolute‐time test}

To capture any common‐mode or non‐$\Delta t$‐dependent correlation, we standardize the pair quantity $w_i$ using the mixed‐event moments:
\begin{eqnarray}
  s_i \;=\;\frac{w_i \;-\;\langle w\rangle_{\rm ME}}
                {\sqrt{\mathrm{Var}_{\rm ME}(w)}},
  \quad
  \langle s\rangle_{\rm ME}=0,\;\langle s^2\rangle_{\rm ME}=1.
\end{eqnarray}
We then form the lifetime product
\begin{eqnarray}
  \tau_i = (t_{1,i}-\bar t)\,(t_{2,i}-\bar t),
  \quad
  \bar t = \langle t\rangle,\;\sigma_t^2 = \mathrm{Var}(t),
\end{eqnarray}
and compute the weighted correlators in SE and ME:
\begin{eqnarray}
  C_\tau^{\rm SE} = \frac{\sum_{i\in\rm SE} s_i\,\tau_i}{\sum_{i\in\rm SE} s_i^2\,\sigma_t^2},
  \quad
  C_\tau^{\rm ME} = \frac{\sum_{j\in\rm ME} s_j\,\tau_j}{\sum_{j\in\rm ME} s_j^2\,\sigma_t^2}.
\end{eqnarray}
The mixed‐event–subtracted statistic
\begin{eqnarray}
  \Delta C_\tau = C_\tau^{\rm SE} - C_\tau^{\rm ME}
\end{eqnarray}
removes acceptance and baseline biases but does not by itself provide the uncertainty or null distribution of \(\Delta C_\tau\).  To estimate its significance, we perform the permutation test using $N_{\rm perm}$ trails (for example, \(N_{\rm perm} \sim10^4\)) in which we randomly shuffle the spin‐weights \(\{s_i\}\) among the lifetime products \(\{\tau_i\}\) in the same‑event sample, recompute
\begin{eqnarray}
  \Delta C_\tau^{(j)}
  = \frac{\sum_i s_i^{(j)}\,\tau_i}{\sum_i (s_i^{(j)})^2\,\sigma_t^2}
    \;-\;
    C_\tau^{\rm ME},
\end{eqnarray}
and build the empirical null distribution of \(\Delta C_\tau\).  The permutation p‑value is then the fraction of trials for which 
\begin{eqnarray}
  \bigl|\Delta C_\tau^{(j)}\bigr|
  \;\ge\;
  \bigl|\Delta C_\tau\bigr|.
\end{eqnarray}
This method captures the combined sampling fluctuations of both \(C_\tau^{\rm SE}\) and its subtraction by \(C_\tau^{\rm ME}\) without relying on any parametric assumptions.

These two complementary, data‑driven tests—a binned $\Delta t$–dependence of the spin coefficient and a permutation‐based absolute‐time correlator—together provide a robust, acceptance‐corrected search for any spin‐dependent departure from independent exponential decay in $\Lambda$–$\bar\Lambda$ pairs.

Additionally, one can also define the simplest lifetime–lifetime covariance $\mathrm{Cov}(t_1, t_2)\;=\; \bigl\langle (t_1 - \bar t)\,(t_2 - \bar t)\bigr\rangle$,  which by construction vanishes when the joint decay probability factorizes as $P_0(t_1,t_2)=P(t_1)P(t_2)$ for independent $\Lambda$–$\bar{\Lambda}$ pairs. Any nonzero value of this covariance—whether arising from spin entanglement or other joint dynamics—would signal genuine time-domain correlations in the decay processes, independent of angular observables. Its simplicity and model‑independence make it a natural first test before moving on to more detailed analyses.  This lifetime–lifetime correlation, if observed, would have profound implications for quantum mechanics. Furthermore, our $\Delta t$‑binned opening‑angle and per‑pair correlator methods extend this baseline test by enabling the detection and characterization of subtle $\Delta t$‑dependent and spin‑weighted features in the entangled decay dynamics.

\vspace{-1ex}

\section{Discussion and Experimental Prospects}

Experiments at BESIII, RHIC, and the LHC routinely reconstruct large numbers of \(\Lambda\!-\!\bar{\Lambda}\) pairs per collision~\cite{BESIII:2025vsr,STAR:2011fbd,ALICE:2013cdo}; for example, in Au+Au collisions at \(\sqrt{s_{NN}}=200\) GeV the STAR experiment observes \(\mathcal{O}(10^2)\) potentially entangled \(\Lambda\!-\!\bar{\Lambda}\) pairings per event and collects \(\sim10^9\) events per year, yielding \(\sim10^{11}\) pairs for analysis.  This sample size, together with well-validated spin-correlation techniques, makes the proposed spin–lifetime and lifetime–lifetime tests experimentally accessible with existing data.  Parameterizing a possible nonfactorizable lifetime covariance as \(\mathrm{Cov}(t_1,t_2)=\epsilon\,\sigma_t^2\), where \(\sigma_t^2=\mathrm{Var}(t)\) and \(\epsilon\ll1\) quantifies the fractional deviation from independent exponential decay, the uncertainty on the sample covariance scales as \(\sigma_t^2/\sqrt{N_{\Lambda\bar\Lambda}}\), so the signal-to-noise ratio is \(\epsilon\sqrt{N_{\Lambda\bar\Lambda}}\).  With \(N_{\Lambda\bar\Lambda}\sim10^{11}\), fractional correlations as small as \(\epsilon\sim10^{-5}\) are within reach at the \(1\sigma\) level (or \(\sim3\times10^{-5}\) for a \(3\sigma\) observation), and even more modest samples would retain sensitivity at the \(\epsilon\sim10^{-3}\) level.  Similarly, the \(\Delta t\)-binned test resolves modulations in \(P(\Delta t)\) at the percent level when \(P\) is extracted in \(K\) bins, since deviations \(\delta P\) become measurable once \(\delta P\gtrsim \mathcal{O}(1)/\sqrt{N_{\Lambda\bar\Lambda}/K}\).  In all cases the mixed-event baseline and permutation trials provide robust control of backgrounds and statistical calibration.

Hyperon production also extends to heavier strange and charm-strange species such as \(\Xi^-\!-\!\bar{\Xi}^+\) and \(\Xi_c\!-\!\bar{\Xi}_c\), and even to mixed-species channels, where analogous lifetime-correlation studies can be performed.  The spin-correlation techniques developed for \(\Lambda\!-\!\bar{\Lambda}\) pairs can be directly adapted, broadening the investigation of time-domain entanglement across baryon families and probing its flavor and mass dependence through hadronization.

The existence of even a small nonzero lifetime covariance would demand a fundamental revision of quantum mechanics’ framework for entanglement, extending the concept beyond its current theoretical boundaries. Furthermore, it would provide fresh insight on how quantum coherence persists in unstable systems~\cite{AlickiLendi2007,Rivas_2012,GardinerZoller2004,Benatti:1998vu,Benatti:2017sjl}
and could serve as a test of wavefunction-collapse models~\cite{Bassi:2012bg,Ghirardi:1985mt,Pearle:1988uh}. Such an observation would further highlight that the conventional separation between decay (as a generalized measurement) and surviving entangled correlations requires refinement when temporal degrees of freedom themselves carry nonclassical structure.  This raises immediate theoretical questions: What mechanisms allow temporal coherence to survive through effectively nonunitary decay processes, and to what extent must existing formulations of decoherence or collapse be extended to accommodate decay-time entanglement?  If a positive signal is observed, possible theoretical responses would include refining effective descriptions of measurement-induced coherence to incorporate correlated decay-time degrees of freedom or generalizing decoherence/collapse frameworks to explicitly allow for entangled temporal observables.  Conversely, a null result would place meaningful upper bounds on any hidden time-domain correlations, sharpening our empirical understanding of entanglement degradation in realistic hadronic environments.  

Because the proposed observables build on established analysis frameworks and do not rely on specific underlying models, they open a new temporal frontier for precision studies of quantum coherence in high-energy collisions.  The combination of immediate feasibility, minimal model dependence, and foundational significance makes these measurements a flexible platform for future extensions and cross-system comparisons.

\section{Conclusion}

We have proposed to test lifetime-lifetime correlations between entangled hyperon pairs. Moreover, we proposed two complementary, data‑driven tests—a binned $\Delta t$ scan of spin–lifetime correlations and a per‑pair lifetime correlator—that together enrich conventional hyperon spin measurements by probing their temporal structure and uncovering subtle entanglement effects.  By reconstructing proper times for each hyperon in the pair, applying mixed‑event baselining, and employing permutation testing, this methodology can be adapted to various baryon species.  Observing a nonzero lifetime–lifetime or spin-lifetime covariance would constitute the first direct evidence of time‑domain entanglement in unstable hadrons, while stringent null results would impose novel constraints on wavefunction‑collapse and decoherence models.

Given the large hyperon yields per collision at BESIII, RHIC and the LHC and the maturity of spin‑correlation frameworks, these measurements can be carried out with existing data sets.  We therefore encourage experimental collaborations to pursue these tests, opening a new frontier in the study of quantum coherence in high‑energy collisions.

\vspace{-1ex}

\section*{Acknowledgement}

We thank Jinhui Chen, Xuguang Huang,  Zhoudunming Tu and Qun Wang for carefully reading the manuscript and providing valuable comments. This work is supported by the US Department of Energy under Grants No. DE-AC02-98CH10886, DE-FG02-89ER40531. 

During the preparation of this work we used OpenAI's latest model o3-mini-high in order to streamline some of the equations and phrasing. After using this tool, we reviewed and edited the content as needed and take full responsibility for the content of the publication.

\bibliographystyle{elsarticle-num-names}
\bibliography{mybib}

\end{document}